\documentstyle[aps,prl]{revtex}

\begin{document}
\draft
\title{ Exact solution of a Kondo lattice model}
\author{Yupeng Wang}
\address{Institute of Physics \& Center for Condensed Matter Physics, 
Chinese Academy of Sciences, 
Beijing 100080, People's Republic of China
}
\maketitle
\begin{abstract}
An integrable Kondo lattice model, which describes a strongly correlated electron host interacting
with a spin-1/2 lattice, is proposed. It is found that with the variations of the Kondo coupling
$J$, the hole concentration $n_h$ and the magnetic field $H$, the system may fall into  a variety of 
phases. The 
phase boundaries of the ground state are determined exactly. The marginal excitations 
and the quantum critical behavior at the phase boundaries are discussed.
\end{abstract}
\pacs{75.20.Hr, 71.10.Hf, 75.30.Mb, 71.27.+a} 

Metallic compounds containing partially filled $f$-bands belong to the general category
of heavy fermions\cite{1}. 
Recently, with the discovery of the Kondo insulators\cite{2} and the non-Fermi-liquid  
behavior,\cite{3}
 the interest in this field has been greatly renewed. Especially, the non-Fermi-liquid behavior
found in some heavy fermion compounds, which stimulates a strong challenge to Landau's Fermi
liquid theory, reveals the quantum critical 
behavior of these systems at low temperatures.\cite{4} The heavy fermion systems are usually modeled 
by the periodic Anderson model
or the Kondo lattice model in some limiting cases. 
 The single-impurity Kondo problem has been studied extensively
and exact solutions\cite{5,6} were obtained. In addition, important progress has been achieved
recently for the Kondo problem in strongly correlated hosts\cite{7,8,9,10}. Nevertheless, 
the understanding to the Kondo lattice
systems is rather unsatisfactory. Though many efforts have been made, an exactly
solvable Kondo lattice model, which may provide us some crucial information of
a heavy fermion system, is still absent. In fact, a generic two-impurity problem is very hard 
to be solved
even in one dimension. I note a few integrable models consisting of many impurities
have been proposed\cite{11,12}. The impurities in these models are very artificial 
and are  almost
independent each other since only forward scattering between the conduction electrons and 
the impurities survive.
The physical effect of these impurities is additive and therefore the problem is still at the
single-impurity level.
\par
In this letter, I propose an exactly solvable Kondo lattice model consisting of
a correlated electron host interacting with a spin-1/2 lattice. In one dimension,  
I show the model is exactly solvable via algebraic Bethe ansatz. The model hamiltonian
reads:
\begin{eqnarray}
H=t\sum_{ i,\delta}h_{i,i+\delta}(1+\vec{\tau}_i
\cdot\vec{\tau}_{i+\delta})
+J\sum_i\vec{S}_i\cdot\vec{\tau}_i,\\
h_{i,j}=\sum_{\sigma=\uparrow,\downarrow}{\cal P}[c_{i,\sigma}^\dagger c_{j,\sigma}
+2(\vec{S}_i\cdot\vec{S}_j+\frac34n_in_j)]{\cal P}\nonumber\\
-(n_i+n_j)+1,\nonumber
\end{eqnarray}
where $c_{i,\sigma}^\dagger$ ($c_{i,\sigma}$) are the creation (annihilation) operators 
of the conduction electrons
with spin $\sigma$ on site $i$; $\vec{S}_i=\frac12\sum_{\sigma,\sigma'}c_{i,\sigma}^\dagger
\vec{\tau}_{\sigma,\sigma'}c_{i,\sigma'}$ are the spin operators of the conduction
electrons ($\vec{\tau}$ the Pauli matrices); $n_i$ denotes the electron number on site $i$;
$\cal {P}$ means the constriction $n_i\leq1$; $\vec{\tau}_i$ are the Pauli matrices 
indicating the local spin on site $i$; $t$ and $J$ are two real constants indicating 
the hoping amplitude and the Kondo coupling constant, respectively; $i=(i_1,\cdots,i_d)$
in $d$ dimension and $\delta$ are the basic vectors of the lattice, $\delta=1$ in
1D and $(1,0)$, $(0,1)$ in 2D etc.. Obviously, Eq.(1) describes an $SU(3)$-invariant $t-J$ 
system\cite{13} coupled with a spin lattice. The first term of Eq.(1) represents the hoping and
interactions of the conduction electrons, which depend not only on the local electron
states, but also on the local spin environment. As discussed in some previous works,\cite{14}
the electron-spin interactions contained in the first term can be mediated either by phonons or
ordinary Coulomb repulsions. The second term describes the usual  Kondo
interactions. \par
In the present form, the solvability of Eq.(1) is rather hidden. We note 
that the first term can be rewritten as $\sum_{i,\delta}P_{i,i+\delta}$, where
$P_{i,j}$ is the permutation operator between the $i$-th site and the $j$-th site.
For any given orthogonal and complete set of Dirac states $\{|\alpha_i>\}$ spanning
the Hilbert space of the $i$-th site, $P_{i,j}$ can be expressed as $\sum_{\alpha,\beta}
X_i^{\alpha\beta}X_j^{\beta\alpha}$, where $X_i^{\alpha\beta}\equiv|\alpha_i><\beta_i|$
represent the Hubbard operators.  Now we check the possible
states of a single site. A natural choice of $\{|\alpha_i>\}$ is $|\gamma_i,\tau_i^z>$,
where $\gamma_i=\uparrow,0,\downarrow$ denote one electron with spin up, no electron and one
electron with spin down, respectively; and $\tau_i^z=\uparrow,\downarrow$ denote
the two components of the local spin. Obviously, the local Hilbert space is 6-dimensional
and therefore the first term of Eq.(1) is $SU(6)$-invariant. To diagonalize the whole
hamiltonian, I introduce the following notations
\begin{eqnarray}
|0>=\frac1{\sqrt2}(|\uparrow,\downarrow>-|\downarrow,\uparrow>),\nonumber\\
|1>=|0,\uparrow>,{~~~}|2>=|0,\downarrow>,{~~~}
|3>=|\uparrow,\uparrow>,\\
|4>=\frac1{\sqrt2}(|\uparrow,\downarrow>+|\downarrow,\uparrow>),
{~~~}|5>=|\downarrow,\downarrow>.\nonumber
\end{eqnarray}
$|0>$ represents a Kondo singlet; $|1>, |2>$
denote two hole states and $|3>, |4>$ and $|5>$ indicate the Kondo triplets. Obviously,
$<\alpha|\beta>=\delta_{\alpha,\beta}$. Therefore, I can rewrite Eq.(1) as (up to
an irrelevant constant)
\begin{eqnarray}
H=2t\sum_{i,\delta}\sum_{\alpha,\beta=0}^5X_i^{\alpha\beta}X_{i+\delta}^{\beta\alpha}
-2J\sum_iX_i^{00}\nonumber\\
-J\sum_i(X_i^{11}+X_i^{22}).
\end{eqnarray}
Now it is clear that the Kondo interaction term represents one of the conserved
quantities of the system. In fact, we have six types of conserved charges
$N_\alpha=\sum_iX_i^{\alpha\alpha}$ which correspond to the number of local state
$|\alpha>$ in the whole system. After this transformation, the hamiltonian (3) takes
 the exact form of an $SU(6)$-invariant
spin chain  model introduced by Sutherland\cite{13} but 
with the Kondo coupling $J$ as an effective field. In any dimensions, we
have five branches of elementary excitations relative to a reference state 
$|\alpha>_g=|\alpha_1>\otimes\cdots\otimes|\alpha_N>$. For example, if we choose $|0>_g$
as the pseudo vacuum, we have two degenerate hole bands (corresponding to
$|0,\uparrow>$ and $|0,\downarrow>$) and three degenerate triplet bands. For the
half-filled case, there is a critical value of the Kondo coupling $J$. When $J>J_c$,
$|0>_g$ becomes the ground state and the degenerate triplet bands are empty. Note
no hole state is allowed in this case. Consider a single excitation upon the ground state
with momentum $\vec {k}=(k_1,\cdots,k_d)$. The excitation energy is $\epsilon(\vec{k})
=4t\sum_{j=1}^d[\cos(k_d)-1]+2J$. Therefore, $J_c=4td$ ($t>0$ is assumed). For $J>J_c$, the triplet
excitations are massive while for $J=J_c$, these excitations are marginal and the system may show
quantum critical behavior. For example, the low temperature specific heat and susceptibility 
at the critical point behave as 
$C\sim T^{d/2}$, $\chi\sim T^{d/2-1}$. In fact, the massive spin excitations in the half filled
case correspond to a Kondo insulator phase if we allow double occupation of electrons on a single site.
This can be realized by replacing the $SU(3)$ $t-J$ term $h_{i,j}$ in Eq.(1) by the $SU(4)$ 
or $SU(2|2)$
Hubbard term.\cite{15} In the latter case, the system exhibits also a charge gap which takes the same value
of the spin gap $\Delta=2(J-J_c)$ at half filling. For any non-half-filling case, we have two 
degenerate
``hole seas" in the ground state configuration and the charge and 
spin excitations are always gapless. However,
there is still a critical point $J_c$. When $J>J_c$, the triplet states are forbidden in the
 ground state. This critical value is generally filling-dependent and is hard to be derived 
 in higher dimensions. The Kondo singlets 
 behave as spin polarons. Because there is a finite gap to
 excite a triplet when $J>J_c$, we expect some type of condensation of the Kondo
 singlets at low temperatures.
 \par
 $Bethe{~}ansatz$. In one dimension, the hamiltonian (3) has the same algebraic structure of
  that introduced by 
 Sutherland\cite{13} but with 
 a very different physical interpretation. Not losing generality, we set $4t=1$ in the following
 text and choose the Kondo insulator state $|0>_g$ 
 as the pseudo vacuum state. The local states $|\alpha_i>$ ($\alpha\geq1$) can be treated as colored
 hard-core bosons with the single occupation condition $\sum_{\alpha=0}^5X_i^{\alpha\alpha}\equiv 1$.
The Bethe ansatz equations (BAE's)  for arbitrary symmetric or supersymmetric models have been given
in Ref.[13]. In our case, they are
 \begin{eqnarray}
 \left(\frac{\lambda^{(1)}_j-\frac i2}{\lambda^{(1)}_j+\frac i2}\right)^N=-\prod_{l=1}^{M_1}
 \frac{\lambda^{(1)}_j-\lambda^{(1)}_l-i}{\lambda^{(1)}_j-\lambda^{(1)}_l+i}\prod_{k=1}^{M_2}
 \frac{\lambda_j^{(1)}-\lambda_k^{(2)}+\frac i2}{\lambda_j^{(1)}-\lambda_k^{(2)}-\frac i2},\\
 \prod_{l=1}^{M_n}\frac{\lambda_j^{(n)}-\lambda_k^{(n)}-i}{\lambda_j^{(n)}-\lambda_k^{(n)}+i}=
 -\prod_{t=n\pm1}\prod_{k=1}^{M_t}\frac{\lambda_j^{(n)}-\lambda_k^{(t)}-\frac i2}
 {\lambda_j^{(n)}-\lambda_k^{(t)}+\frac i2},\nonumber\\
 n=2,3,4,5;\nonumber
 \end{eqnarray}
 with the eigenvalue of the hamiltonian (3) as 
 \begin{eqnarray}
 E=-\sum_{j=1}^{M_1}(\frac {\frac12}{{\lambda_j^{(1)}}^2+\frac14}-2J)-J(N_1+N_2),
 \end{eqnarray}
 where $M_n=N_n+\cdots+N_5$ and $M_6\equiv 0$, $\lambda_j^{(n)}$ are the rapidities of the flavor 
 waves (holons and spinons).
 Notice that the boundary condition $X_1^{\alpha\beta}=X_{N+1}^{\alpha\beta}$ has been used in deriving
 Eq.(4) and an irrelevant constant has been omitted in Eq.(5). Relative to the Kondo insulator 
 state $|0>_g$, the unoccupied states $|0,\uparrow>$ and $|0,\downarrow>$
 can be treated as Kondo holes which are responsible to the dynamical properties of a doped Kondo insulator.
 For large enough $J>J_c$, there is no triplet states in the ground state configuration, i.e., $M_3=0$.
 The effective low-energy hamiltonian of the system is therefore equivalent to an $SU(3)$-invariant
 $t-J$ model.
 \par
 $The{~} phase{~} boundary{~}J_c(n_h)$. For a given Kondo-hole concentration $n_h=(N_1+N_2)/N$, 
 there is a phase boundary $J_c(n_h)$ above which the spin triplet states will
 be eliminated from the ground state.  Set $\rho_n(\lambda)$ as the 
 distribution of $\lambda^{(n)}$-modes in the ground state. When $J>J_c$, from Eq.(4) we have
 \begin{eqnarray}
 \rho_1(\lambda)=a_1(\lambda)+\int_{-\Lambda_2}^{\Lambda_2}a_1(\lambda-\nu)\rho_2(\nu)
 d\nu-\int_{-\Lambda_1}^{\Lambda_1}a_2(\lambda-\nu)\rho_1(\nu)d\nu,\nonumber\\
 \rho_2(\lambda)=\int_{-\Lambda_1}^{\Lambda_1}a_1(\lambda-\nu)\rho_1(\nu)d\nu
 -\int_{-\Lambda_2}^{\Lambda_2}a_2(\lambda-\nu)\rho_2(\nu)d\nu,\\
 \rho_n(\lambda)=0,{~~~~~~~~~~}n\geq3,\nonumber
 \end{eqnarray}
 where $a_n(\lambda)=n/[2\pi(\lambda^2+n^2/4)]$ and the cutoffs $\Lambda_{1,2}$ are determined by
 \begin{eqnarray}
 \int_{-\Lambda_1}^{\Lambda_1}\rho_1(\lambda)d\lambda=n_h,{~~~~}
 \int_{-\Lambda_2}^{\Lambda_2}\rho_1(\lambda)d\lambda=\frac 12n_h.
 \end{eqnarray}
 By integrating the second equation of Eq.(6), we obtain $\Lambda_2=\infty$. The excitation energy 
 of a triplet mode can be exactly derived 
 by considering the process $M_1\to M_1+1$, $M_2\to M_2+1$ and $M_3=1$. Such an excitation 
 can be realized by adding a $\lambda^{(1)}$
 mode $\lambda_p^{(1)}$ above the $\lambda^{(1)}$-sea, a $\lambda^{(3)}$-mode $\lambda^{(3)}_p$ and a 
 $\lambda^{(2)}$ hole
 $\lambda_h^{(2)}$ in the $\lambda^{(2)}$-sea. After some manipulation we obtain the excitation energy 
 $\Delta E(\lambda_p^{(1)},\lambda_h^{(2)},\lambda^{(3)}_p)$ as
 \begin{eqnarray}
 \Delta E=\epsilon_1(\lambda_p^{(1)})-\epsilon_2(\lambda_h^{(2)})+\epsilon_3(\lambda^{(3)}_p),
 \end{eqnarray}
 where the dressed energy $\epsilon_n(\lambda)$ satisfy (see, for example, ref.[16])
 \begin{eqnarray}
 \epsilon_1(\lambda)=-\pi a_1(\lambda)-\mu+\int_{-\infty}^\infty a_1(\lambda-\nu)\epsilon_2(\nu)
 d\nu\nonumber\\-\int_{-\Lambda_1}^{\Lambda_1}a_2(\lambda-\nu)\epsilon_1(\nu)d\nu,\nonumber\\
 \epsilon_2(\lambda)=\int_{-\Lambda_1}^{\Lambda_1}a_1(\lambda-\nu)\epsilon_1(\nu)d\nu
 -\int_{-\infty}^\infty a_2(\lambda-\nu)\epsilon_2(\nu)d\nu,\\
 \epsilon_3(\lambda)=\mu+2J+\int_{-\infty}^\infty a_1(\lambda-\nu)\epsilon_2(\nu)d\nu,\nonumber
 \end{eqnarray}
 where $\mu=-\pi a_1(\Lambda_1)$ denotes the chemical potential. The energy gap associated with this
 excitation is given by $\Delta(n_h)=\epsilon_3(0)$ (Notice that $\epsilon_1(\pm\Lambda_1)=0$ and
 $\epsilon_2(\pm\infty)=0$). $\Delta$ is a monotonically increasing function of $n_h$ and ranges from
 $2(J-2)$ for $n_h=0$ to $2J$ for $n_h=1$, while $J_c$ is a monotonically decreasing function of
 $n_h$ and ranges from $J_c=2$ for $n_h=0$ to $J_c=0$ for $n_h=1$. For $n_h=2/3$, the dressed energy
 $\epsilon_3(\lambda)$ reads:
 \begin{eqnarray}
 \epsilon_3(\lambda,n_h=2/3)=-\frac12\int\frac{e^{-\frac12|\omega|}e^{-i\omega\lambda}}{4\cosh^2\frac
 \omega2-1}d\omega+2J.
 \end{eqnarray}
 The energy gap $\Delta(2/3)$ and the critical value $J_c(2/3)$ can be easily derived as
 \begin{eqnarray}
 \Delta(2/3)=2J-\frac\pi{2\sqrt3}+\frac12\ln3,{~~~~~~}
 J_c(2/3)=\frac\pi{4\sqrt3}-\frac14\ln3.
 \end{eqnarray}
 \par
 $Quantum{~} critical{~} behavior{~} at{~} J=J_c(n_h)$. When $J<J_c(n_h)$, the triplet excitations 
 become massless and the system behaves as a Luttinger liquid
 with a holon band and four spinon bands. Exactly at the critical point $J=J_c(n_h)$, the triplet 
 excitations 
 are marginal, indicating a quantum phase transition at this point. To see it clearly, 
 let us consider the dispersion relation of the triplet excitations for $J=J_c(n_h)$.
 From the third equation of the BAE's (4) we know that a single $\lambda^{(3)}$-mode with rapidity 
 $\lambda$ is quantized as
 \begin{eqnarray}
 \frac{2\pi I}N=\frac1N\sum_{j=1}^{M_2}2\arctan[2(\lambda-\lambda_j^{(2)})],
 \end{eqnarray}
 where $I$ is an arbitrary integer or half integer depending on the parity of $M_2$.
  Therefore, the left hand side of Eq.(12) can be treated 
 as the quasi-momenta $k(\lambda)$ of the $\lambda^{(3)}$-mode.\cite{16}
 In the thermodynamic limit $N\to\infty$,
 \begin{eqnarray}
 k(\lambda)=2\int_{-\infty}^{\infty}\arctan[2(\lambda-\nu)]\rho_2(\nu)d\nu.
 \end{eqnarray}
 From Eqs.(6) and (9) I find that the velocity of the $\lambda^{(3)}$ 
 mode $v_3=\lim_{k\to0}
 \partial\epsilon_3(\lambda)/\partial k(\lambda)=0$, implying a finite mass of this excitation. Therefore,
 the dispersion relation takes the form: $\epsilon_3[k(\lambda)]=k^2/(2m)$, for $k\to0$. The effective mass 
 of the excitation reads
 \begin{eqnarray}
 m=\lim_{\lambda\to0}\left[\frac{\partial^2\epsilon_3(\lambda)}{\partial k^2(\lambda)}\right]^{-1}.
 \end{eqnarray}
 It can be easily demonstrated that $m$ takes positive values for any given $n_h$.
 At very low temperatures, only few excitations exist and behave as a quasi-ideal quantum gas
 obeying the Pauli exclusion principle but with a zero effective chemical potential.
 The quantum critical behavior at $J=J_c(n_h)$ is mainly governed by the marginal excitations. For
 example, the low-temperature specific heat and susceptibility of the system behave as
 \begin{eqnarray}
 C\sim T^{\frac12},{~~~~~}\chi\sim T^{-\frac12}.
 \end{eqnarray}
 The divergence of the susceptibility at $T=0$ is due to the singularity of the density of states of
 the marginal excitations.
 \par
 $Response{~} to{~} the{~} magnetic{~} field$. When $J<J_c(n_h)$, the system behaves as a five-component
 Luttinger liquid, while for $J>J_c(n_h)$, an external magnetic field may drive some quantum phase
 transitions at zero temperature. For a given hole concentration
 $n_h$, there are three critical fields $H_c^1$, $H_c^2$ and $H_c^3$. In a very weak field,
 the susceptibility is Pauli type due to the response of the Kondo holes.
 When $H$ reaches $H_c^1$, the Kondo holes
 are completely polarized, while when $H=H_c^2$, the Kondo singlets begin to be polarized and the
 zero-temperature susceptibility has a singularity at this point, $\chi(H)\sim (H-H_c^2)^{-\frac12}$ 
 ($H\geq H_c^2$). For a strong enough $H\geq H_c^3$, both the Kondo holes and the spin singlets are 
 completely polarized. $H_c^1$ ranges from $0$ ($n_h=0$) to $4$ ($n_h
 =1$), $2(J-2)\leq H_c^2\leq 2J$  and $H_c^3$ ranges from $2(J+2)$ ($n_h=0$)
 to $4$ ($n_h=1$).  For $H_c^1<H<H_c^2$, 
 a magnetization plateau occurs since in this case, the Kondo holes are completely polarized 
 while $H$ is still not strong
 enough to excite the triplet modes. For $H_c^2<H_c^1$, singlet polarization
 occurs before the saturation of the holes' magnetization. There is no magnetization plateau for
 $H<H_c^3$ but there are a singularity at $H=H_c^2$ and a kink at $H=H_c^1$ in the $\chi-H$ curve.
 To see the situation clearly, let us consider 
 the ground state properties of $n_h=1/2$ and $J>J_c(1/2)$ case. With a magnetic field, the eigenenergy reads (up to
 an irrelevant constant)
 \begin{eqnarray}
 E=-\sum_{j=1}^{M_1}\frac{\frac12}{{\lambda_j^{(1)}}^2+\frac14}-2JN_0
 -H(N_3-N_5)-\frac12H(N_1-N_2).
 \end{eqnarray}
 In this case, $H_c^1<H_c^2<H_c^3$. When $H=H_c^1$, the ground state configuration is described 
 by $N_0=N/2=N_1=M_1$
 and $N_n=0$ for $n>2$. The density of $\lambda^{(1)}$ is still given by Eq.(6) but with $\rho_2(\lambda)=0$
 and $\Lambda_1=\infty$. The critical field $H_c^1$ can be derived by considering the excitation process
 $N_1\to N/2-1$, $N_2\to1$. This is realized by putting a hole in the $\lambda^{(1)}$ sea and adding
 a particle to the $\lambda^{(2)}$ band. Denoting the rapidities of the hole and the particle as 
 $\lambda_h$ and $\lambda_p$, respectively and setting $\delta\rho_1(\lambda)/N$ as the change of 
 $\rho_1(\lambda)$ due to $\lambda_h$ and $\lambda_p$, from Eq.(6) we have
 \begin{eqnarray}
 \delta\rho_1(\lambda)=-\int_{-\infty}^\infty a_2(\lambda-\nu)\delta\rho_1(\nu)d\nu
 \nonumber\\+a_1(\lambda-\lambda_p)
 -\delta(\lambda-\lambda_h).
 \end{eqnarray}
 The excitation energy associated with this excitation is
 \begin{eqnarray}
 \epsilon(\lambda_h,\lambda_p)=-\pi\int_{-\infty}^\infty a_1(\lambda)\delta\rho_1(\lambda)d\lambda+H.
 \end{eqnarray}
 Solving Eq.(17) by Fourier transformation and substituting it into Eq.(18), we readily obtain 
 the energy gap
 $\Delta_1=\epsilon(\infty,0)=H-\ln2$. Obviously, $H_c^1=\ln2$. $H_c^2$ can be derived 
 in a similar way. 
 For convenience, we choose $|1>_g$ as the vacuum state. The excitation breaking a Kondo singlet corresponds 
 to $M_1=N_0\to N/2-1$, $M_2\to 1$. The energy gap associated with this excitation is 
$\Delta_2=2J-H-\ln2$. Therefore, $H_c^2=2J-\ln2$. To derive $H_c^3$, we choose still $|1>_g$ 
as the vacuum state. When $H>H_c^3$, the ground state is described by $N_1=N_3=N/2$. With the
same procedure we readily obtain the energy gap associated with this excitation reads $\Delta_3=H-2J-\ln2$.
 Therefore
$H_c^3=2J+\ln2$. 
 \par
 $Thermal{~} BAE's$. 
 The thermal BAE's  can be derived by following the 
 standard method. \cite{17,18} Here I give the result and omit the details.
 After some manipulation, we obtain the density of the free energy as
 \begin{eqnarray}
 f=-T\sum_{n=1}^\infty\int a_n(\lambda)\ln[1+\eta_{n,1}^{-1}(\lambda)]d\lambda,
 \end{eqnarray}
 where $\eta_{n,\alpha}(\lambda)$ are the elements of the following integral equations
 \begin{eqnarray}
 G[\ln(1+\eta_{n+1,\alpha})+(1-\delta_{n,1})\ln(1+\eta_{n-1,\alpha})]\nonumber\\
 =\frac {\pi g(\lambda)}T\delta_{n,1}\delta_{\alpha,1}+G[(1-\delta_{\alpha,5})\ln(1+\eta_{n,\alpha+1}^{-1})
 \nonumber\\+(1-\delta_{\alpha,1})\ln(1+\eta_{n,\alpha-1}^{-1})]+\ln\eta_{n,\alpha},\\
 n=1,2\cdots;{~~~~~~~}\alpha=1,2\cdots 5,\nonumber
 \end{eqnarray}
with the boundary conditions
\begin{eqnarray}
\lim_{n\to\infty}\frac{\ln \eta_{n,\alpha}}n=\zeta_\alpha,
\end{eqnarray}
where $G$ is an integral operator with the kernel $g(\lambda)=1/2\cosh(\pi\lambda)$; $\zeta_1=J/T-\mu/T-H/2T$,
$\zeta_2=\zeta_4=\zeta_5=H/T$, $\zeta_3=J/T+\mu/T-3H/2$ and $\mu$ denotes the chemical potential of the Kondo holes.
Eqs.(20-21) can be solved with asymptotic expansions\cite{5,6} or numerically. Details will be given elsewhere.
\par
The author is  indebted to the hospitality
of Institut f\"{u}r Physik, Universit\"{a}t Augsburg. This work was partially supported by
AvH-Stiftung, NSFC, FCAS and NOYSFC.


\begin{references}
\bibitem{1}For a review, see, P.A. Lee, T.M. Rice, J.W. Serene, L.J. Sham and J.W. Wilkins, Comments Cond. Matt.
Phys. {\bf 12}, 99 (1986).
\bibitem{2}For a review, see, G. Aeppli and Z. Fisk, Comments Cond. Matt. Phys. {\bf 16}, 155 (1992).
\bibitem{3}See, for example,  F. Steglich et al., J. Phys: Cond. Matt.
{\bf 8}, 9909 (1996); H. von L\"{o}hneysen, {\bf ibid}, 9689 (1996) and references therein.
\bibitem{4}N.D. Mathur, F.M. Grosche, S.R. Julian, I.R. Walker, D.M. Freye, R.K.W. Haselwimmer, and
G.G. Lonzarich, Nature {\bf 394}, 39 (1998).
\bibitem{5}N. Andrei, K. Furuya and J. Lowenstein, Rev. Mod. Phys. {\bf 55}, 331 (1983).
\bibitem{6}A.M. Tsvelik and P.B. Wiegmann, Adv. Phys. {\bf 32}, 453 (1983).
\bibitem{7}D.-H. Lee and J. Toner, Phys. Rev. Lett. {\bf 69}, 3378 (1992).
\bibitem{8}D. Withoff and E. Fradkin, Phys. Rev. Lett. {\bf 64}, 1835 (1990).
\bibitem{9}A. Furusaki and N. Nagaosa, Phys. Rev. Lett. {\bf 72}, 892 (1994).
\bibitem{10}Y. Wang, J.H. Dai, Z.N. Hu and F.C. Pu, Phys. Rev. Lett. {\bf 79}, 1901 (1997).
\bibitem{11}See, for example, P. Schlottmann, Phys. Rev. B {\bf 57}, 10638 (1998) and references therein.
\bibitem{12}Y. Wang, J. Dai, F.-C. Pu and U. Eckern, Phys. Rev. B {\bf 59}, 7393 (1999).
\bibitem{13}B. Sutherland, Phys. Rev. B {\bf 12}, 3795 (1975).
\bibitem{14}A.A. Nersesyan and A.M. Tsvelik, Phys. Rev. Lett. {\bf 78}, 3939 (1997);
D.G. Shelton and A.M. Tsvelik, Phys. Rev. B {\bf 53}, 14036 (1993).
\bibitem{15}F.H.L. Essler, V.E. Korepin and K. Shoutens, Phys. Rev. Lett. {\bf 68}, 2960 (1992).
\bibitem{16}For example, see, V.E. Korepin, A.G. Izergin and N.M. Bogoliubov, Quantum Inverse Scattering Method, Correlation
Functions and Algebraic Bethe Ansatz (Cambridge University Press, Cambridge 1993).
\bibitem{17}C.N. Yang and C.P. Yang, J. Math. Phys. {\bf 10}, 1115 (1969).
\bibitem{18}M. Takahashi, Prog. Theor. Phys. {\bf 46}, 401; 1388 (1971). 



\end{references}
\end{document}